\pdfoutput=1
\documentclass[12pt,a4paper]{article}

\usepackage{lineno}  

\usepackage{graphicx}  

\usepackage{xspace}
\usepackage{color}
\usepackage{colortbl}
\usepackage{subfigure}
\usepackage{amsmath}

\usepackage{ifthen} 
\newboolean{pdflatex}
\setboolean{pdflatex}{true} 
\usepackage{rotating} 
\newboolean{articletitles}
\setboolean{articletitles}{true} 

\newboolean{uprightparticles}
\setboolean{uprightparticles}{false} 
\usepackage{amssymb}
\usepackage{amsfonts}
\usepackage{upgreek}
\usepackage{lhcb-symbols-def}
\usepackage{hyperref}    
\usepackage[all]{hypcap} 

\usepackage{bm}
\usepackage{afterpage}

\usepackage{epsfig,accents}

\newcommand*{\fancybar}{\scalebox{.4}{(}\raisebox{-1.7pt}{-}\scalebox{.4}{)}}
\newcommand*{\brabar}[1]{\accentset{\fancybar}{#1}}

\newcommand{\bea}{\begin{eqnarray}}
\newcommand{\eea}{\end{eqnarray}}
\newcommand{\beq}{\begin{equation}}
\newcommand{\eeq}{\end{equation}}

\usepackage{cite}
\usepackage{mciteplus}
\begin{document}
\renewcommand{\thefootnote}{\fnsymbol{footnote}}
\setcounter{footnote}{1}



\begin{titlepage}
\pagenumbering{roman}
\belowpdfbookmark{Title page}{title}

\pagenumbering{roman}
\vspace*{-1.5cm}
\centerline{\large EUROPEAN ORGANIZATION FOR NUCLEAR RESEARCH (CERN)}
\vspace*{1.5cm}
\hspace*{-5mm}\begin{tabular*}{16cm}{lc@{\extracolsep{\fill}}r}
 & & CERN-PH-EP-2012-107\\
 & & LHCb-PAPER-2012-006\\  
 & &April 25, 2012 \\ 
 & & \\
\end{tabular*}

\vspace*{3.0cm}

{\bf\boldmath\Large
\begin{center}
Measurement of the \CP-violating phase $\phi_s$ in $\Bsb\to \jpsi \pi^+\pi^-$ decays\\
\end{center}
}

\vspace*{2.0cm}
\begin{center}
\normalsize {
The LHCb collaboration\footnote{Authors are listed on the following pages.}
}
\end{center}

\begin{abstract}
  \noindent
Measurement of the mixing-induced \CP-violating phase $\phi_s$ in $\Bsb$ decays is of prime importance in probing new physics. Here 7421$\pm$105 signal events from the dominantly \CP-odd final state $\jpsi\pi^+\pi^-$ are selected in 1\,fb$^{-1}$ of $pp$ collision data collected at $\sqrt{s} = 7$~TeV with the LHCb detector. A time-dependent fit to the data yields a value of $\phi_s=-0.019^{+0.173+0.004}_{-0.174-0.003}$\,rad, consistent with the Standard Model expectation. No evidence of direct \CP violation is found.

\end{abstract}

\vspace*{2.0cm}
\vspace{\fill}

\vspace*{1.0cm}
\vspace{\fill}

\vspace*{1.0cm}
\hspace*{6mm}Submitted to Physics Letters B\\
\newpage
\begin{center}
{\bf LHCb collaboration}\\
\end{center}
\begin{flushleft}
R.~Aaij$^{38}$, 
C.~Abellan~Beteta$^{33,n}$, 
B.~Adeva$^{34}$, 
M.~Adinolfi$^{43}$, 
C.~Adrover$^{6}$, 
A.~Affolder$^{49}$, 
Z.~Ajaltouni$^{5}$, 
J.~Albrecht$^{35}$, 
F.~Alessio$^{35}$, 
M.~Alexander$^{48}$
% ...

{\footnotesize \it
$ ^{1}$Centro Brasileiro de Pesquisas F\'{i}sicas (CBPF), Rio de Janeiro, Brazil\\
% ...
$ ^{5}$Clermont Universit\'{e}, Universit\'{e} Blaise Pascal, CNRS/IN2P3, LPC, Clermont-Ferrand, France\\
$ ^{6}$CPPM, Aix-Marseille Universit\'{e}, CNRS/IN2P3, Marseille, France\\
% ...
$ ^{33}$Universitat de Barcelona, Barcelona, Spain\\
$ ^{34}$Universidad de Santiago de Compostela, Santiago de Compostela, Spain\\
$ ^{35}$European Organization for Nuclear Research (CERN), Geneva, Switzerland\\
% ...
$ ^{38}$Nikhef National Institute for Subatomic Physics, Amsterdam, The Netherlands\\
% ...
$ ^{43}$H.H. Wills Physics Laboratory, University of Bristol, Bristol, United Kingdom\\
% ...
$ ^{48}$School of Physics and Astronomy, University of Glasgow, Glasgow, United Kingdom\\
$ ^{49}$Oliver Lodge Laboratory, University of Liverpool, Liverpool, United Kingdom\\
% ...
$ ^{n}$LIFAELS, La Salle, Universitat Ramon Llull, Barcelona, Spain\\
% ...
}
\end{flushleft}

\end{titlepage}
\renewcommand{\thefootnote}{\arabic{footnote}}
\setcounter{footnote}{0}

\pagestyle{empty}  

\pagestyle{plain} 
\setcounter{page}{1}
\pagenumbering{arabic}


%



\section{Introduction}

Current knowledge of the Cabibbo-Kobayashi-Maskawa (CKM) matrix leads to the Standard Model (SM) expectation that the mixing-induced \CP violation phase in $\Bsb$ decays proceeding via the $b\to c\overline{c}s$ transition is small and accurately predicted \cite{Charles:2011va}.
Therefore, new physics can be decisively revealed by its measurement. This phase denoted by $\phi_s$ is given in the SM by $-2\arg\left[{V_{ts}V_{tb}^*}/{V_{cs}V_{cb}^*}\right]$, where the $V_{ij}$ are elements of the CKM matrix. 
Motivated by a prediction in Ref.~\cite{Stone:2008ak}, the LHCb collaboration made the first observation of $\Bsb\to \jpsi f_0(980)$,  $f_0(980)\to\pi^+\pi^-$ \cite{Aaij:2011fx}, which was subsequently confirmed by others \cite{Li:2011pg,*Abazov:2011hv,Aaltonen:2011nk}. This mode is a \CP-odd eigenstate and its use obviates the need to perform an angular analysis in order to determine $\phi_s$ \cite{LHCb:2011ab}, as is required in the $\jpsi\phi$ final state \cite{LHCb:2011aa,Abazov:2011ry,*CDF:2011af}. 
In this Letter we measure $\phi_s$ using the final state $\jpsi\pi^+\pi^-$ over a large range of $\pi^+\pi^-$ masses, 775$-$1550 MeV,\footnote{We work in units where $c=\hbar=1$.} which has been shown to be an almost pure \CP-odd eigenstate \cite{LHCb-PAPER-2012-005}. We designate events in this region as $f_{\rm odd}$.
This phase is the same as that measured in $\jpsi \phi$ decays, ignoring contributions from suppressed processes \cite{Faller:2008gt,*Fleischer:2011au}.

The decay time evolutions for initial $B_s^0$ and $\Bsb$ decaying into a \CP-odd eigenstate, $f_-$, assuming only one CKM phase, are \cite{Nierste:2009wg,*Bigi:2000yz}
\begin{equation}
   \label{eq:CPrate}
\Gamma\left(
\brabar{B}_s^0\to f_-\right) = {\cal N}e^{-\Gamma_s t}\, \Bigg\{ \frac{e^{\Delta\Gamma_s t/2}}{2}(1+\cos\phi_s)+ \frac{e^{-\Delta\Gamma_s t/2}}{2}(1-\cos\phi_s) \pm\sin\phi_s\sin \left( \dm_s \, t \right) \Bigg\}\,, 
\end{equation}
where $\Delta\Gamma_s=\Gamma_{\rm L}-\Gamma_{\rm H}$ is the decay width difference between light and heavy mass eigenstates, $\Gamma_s=(\Gamma_{\rm L}+\Gamma_{\rm H})/2$ is the average decay width, $\Delta m_s=m_{\rm H}-m_{\rm L}$ is the mass difference, and $ {\cal N}$ is a time-independent normalization factor. The plus sign in front of the $\sin\phi_s$ term applies to an initial $\Bsb$ and the minus sign to an initial $B^0_s$ meson.
The time evolution of the untagged rate is then
\begin{equation}
\Gamma\left( B_s^0\to f_-\right) +  \Gamma\left( \Bsb\to f_-\right)=
{ \cal N}e^{-\Gamma_s t}\, \Bigg\{e^{\Delta\Gamma_s t/2}(1+\cos\phi_s)+ e^{-\Delta\Gamma_s t/2}(1-\cos\phi_s)\Bigg\}\,.
\label{eq:untagged}
\end{equation}
Note that there is information in the shape of the lifetime distribution that correlates $\Delta\Gamma_s$ and $\phi_s$.
In this analysis we will use samples of both flavour tagged and untagged decays. 
Both Eqs.~\ref{eq:CPrate} and \ref{eq:untagged} are invariant under the change $\phi_s\to \pi-\phi_s$ when $\Delta\Gamma_s\to -\Delta\Gamma_s$, which gives an inherent ambiguity. Recently this ambiguity has been resolved \cite{Aaij:2012eq}, so only the allowed solution with $\Delta\Gamma_s>0$ will be considered.

\section{Data sample and selection requirements}
The data sample consists of 1\,fb$^{-1}$ of integrated luminosity collected with the \lhcb detector \cite{Alves:2008zz} at 7~TeV  centre-of-mass energy in $pp$ collisions at the LHC.  
The detector is a single-arm forward
spectrometer covering the pseudorapidity range $2<\eta <5$, designed
for the study of particles containing \bquark or \cquark quarks. Components include a high-precision tracking system consisting of a
silicon-strip vertex detector surrounding the $pp$ interaction region,
a large-area silicon-strip detector located upstream of a dipole
magnet with a bending power of about $4{\rm\,Tm}$, and three stations
of silicon-strip detectors and straw drift-tubes placed
downstream. The combined tracking system has a momentum resolution
$\delta p/p$ that varies from 0.4\% at 5\gev to 0.6\% at 100\gev,
and an impact parameter (IP) resolution of 20\mum for tracks with high
transverse momentum (\pt).  Charged hadrons are identified using two
ring-imaging Cherenkov (RICH) detectors. Photon, electron and hadron
candidates are identified by a calorimeter system consisting of
scintillating-pad and pre-shower detectors, an electromagnetic
calorimeter and a hadronic calorimeter. Muons are identified by a muon
system composed of alternating layers of iron and multiwire
proportional chambers. The trigger consists of a hardware stage, based
on information from the calorimeter and muon systems, followed by a
software stage which applies a full event reconstruction. 

Events were triggered by detecting two muons with an invariant mass within 120\, MeV of the nominal  \jpsi mass \cite{PDG}. To be considered 
a $\jpsi$ candidate, particles of opposite charge are required to have  $p_{\rm T}$ greater than 500\,MeV, be identified as muons, and 
form a vertex with fit  $\chi^2$ per number of degrees of freedom less than 16. Only candidates with  a dimuon invariant mass between $-$48~MeV and +43 MeV of the $\jpsi$ mass peak are selected. For further analysis the four-momenta of the dimuons are constrained to yield the $\jpsi$ mass.

For this analysis we use a Boosted Decision Tree (BDT) \cite{Hocker:2007ht} to set the $\jpsi\pi^+\pi^-$ selection requirements. We first implement a preselection that preserves a large fraction of the signal events,
including the requirements that the pions have \pt $>$ 250\,MeV and  be identified by the RICH. \Bsb candidate decay tracks must form a common vertex that is detached from the primary vertex. The angle between the combined momentum vector of the decay products and
the vector formed from the positions of the primary and the \Bsb decay vertices (pointing angle)  is required to be consistent with zero. If more
than one primary vertex is found the one corresponding to the smallest IP significance of the $\Bsb$ candidate is chosen.

The variables used in the BDT are the muon identification quality, the probability that the $\pi^{\pm}$ come from the primary vertex  (implemented in terms of the  IP \chisq), the  \pt of each pion, the $\Bsb$ vertex $\chi^2$, the pointing angle and the $\Bsb$ flight distance from production to decay vertex. For various calibrations we also analyze samples of $\Bzb\to \jpsi \Kstarzb$, $\Kstarzb\to\pi^+K^-$, and its charge-conjugate. The same selections are used as for $\jpsi\pi^+\pi^-$ except for particle identification.

The BDT is trained with $\Bsb\to \jpsi f_0(980)$ Monte Carlo events generated using \pythia \cite{Sjostrand:2006za} and the LHCb detector simulation based on G{\sc eant}4 \cite{Agostinelli:2002hh}.
The following two data samples are used to study the background. The first contains $\jpsi \pi^+\pi^+$ and $\jpsi \pi^-\pi^-$ events 
with $m(\jpsi \pi^{\pm}\pi^{\pm})$  within $\pm$50 MeV of the \Bsb mass, called the like-sign sample. The second consists of 
events in the $\Bsb$ sideband having $m(\jpsi\pi^+\pi^-)$ between 200 and 250 MeV above the $\Bsb$ mass peak.  In both cases we require $775 < m(\pi\pi)<$~1550\,MeV.

\begin{figure}[hbt]
\centering
\includegraphics[width=4.in]{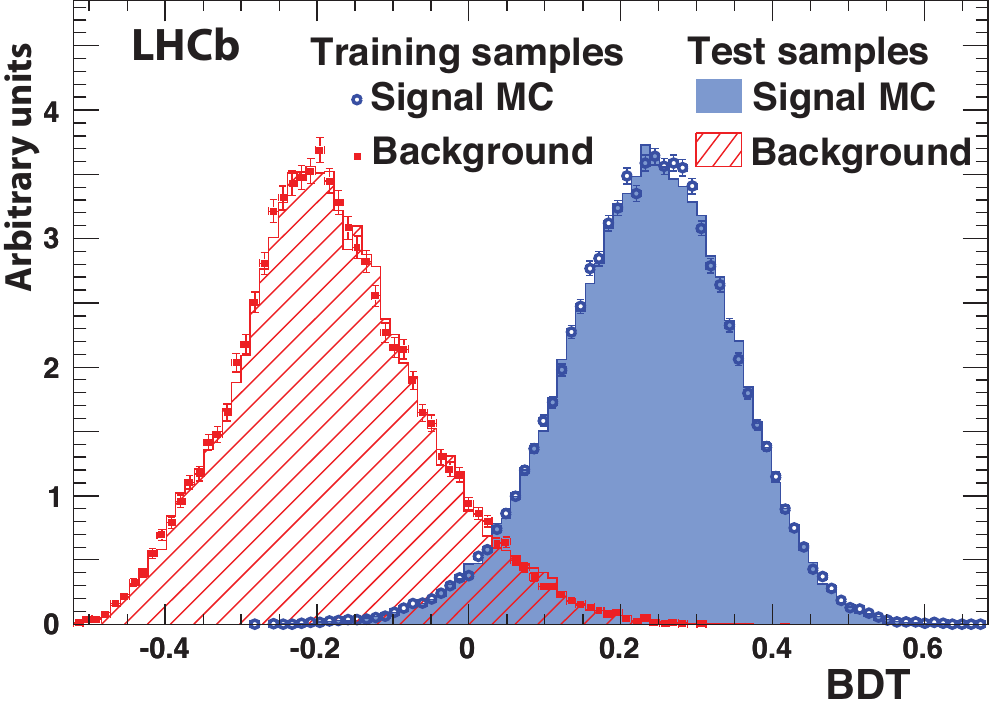}
\vspace{-3mm}
\caption{
Distributions of the BDT variable for both training and test samples of $\jpsi\pi\pi$ signal and background events.
The signal samples are from simulation and the background samples derived from data.
 } \label{overtrain_BDT}
\end{figure}

Separate samples are used to train and test the BDT. Training samples consist of 74,230 signal  and 31,508 background events, while the testing samples contain 74,100 signal and 21,100 background events. 
Figure~\ref{overtrain_BDT} shows the signal and background BDT distributions of the training and test samples. The training and test samples are in excellent agreement.  We select $\Bsb\to \jpsi\pi^+\pi^-$ candidates with BDT $>$ 0 to maximize signal significance for further analysis. 

\begin{figure}[hbt]
\centering
\vspace{-4mm}
\includegraphics[width=4.5in]{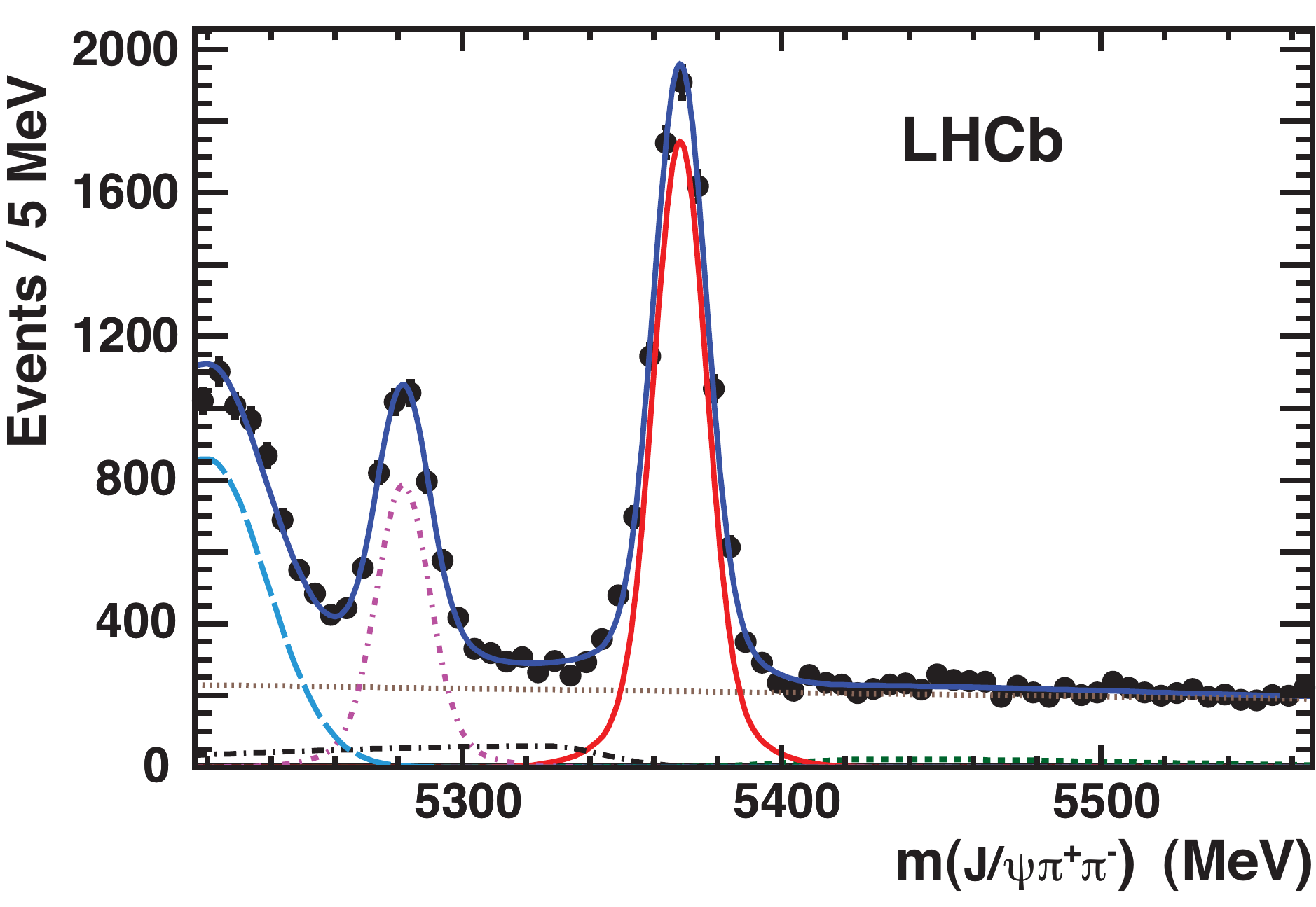}
\vspace{-4mm}
\caption{Mass distribution of the selected $\jpsi \pi^+\pi^-$ combinations in the $f_{\rm odd}$ region. The blue solid curve shows the result of a fit with a double Gaussian signal (red solid curve) and several background components:  combinatorial background (brown dotted line),  background from $B^-\to\jpsi K^-$ and $\jpsi \pi^-$ (green short-dashed line), $\Bzb\rightarrow \jpsi \pi^+\pi^-$ (purple dot-dashed),   $\Bsb\rightarrow \jpsi\eta'$ and $\Bsb\rightarrow \jpsi\phi$ when  $\phi\to\pi^+\pi^-\pi^0$ (black dot-long-dashed), and  $\Bdb\rightarrow \jpsi K^- \pi^+$ (light-blue long-dashed). 
 } \label{fitmass_bdt_sel}
\end{figure}

\begin{figure}[h!]
\centering
\includegraphics[width=4.5in]{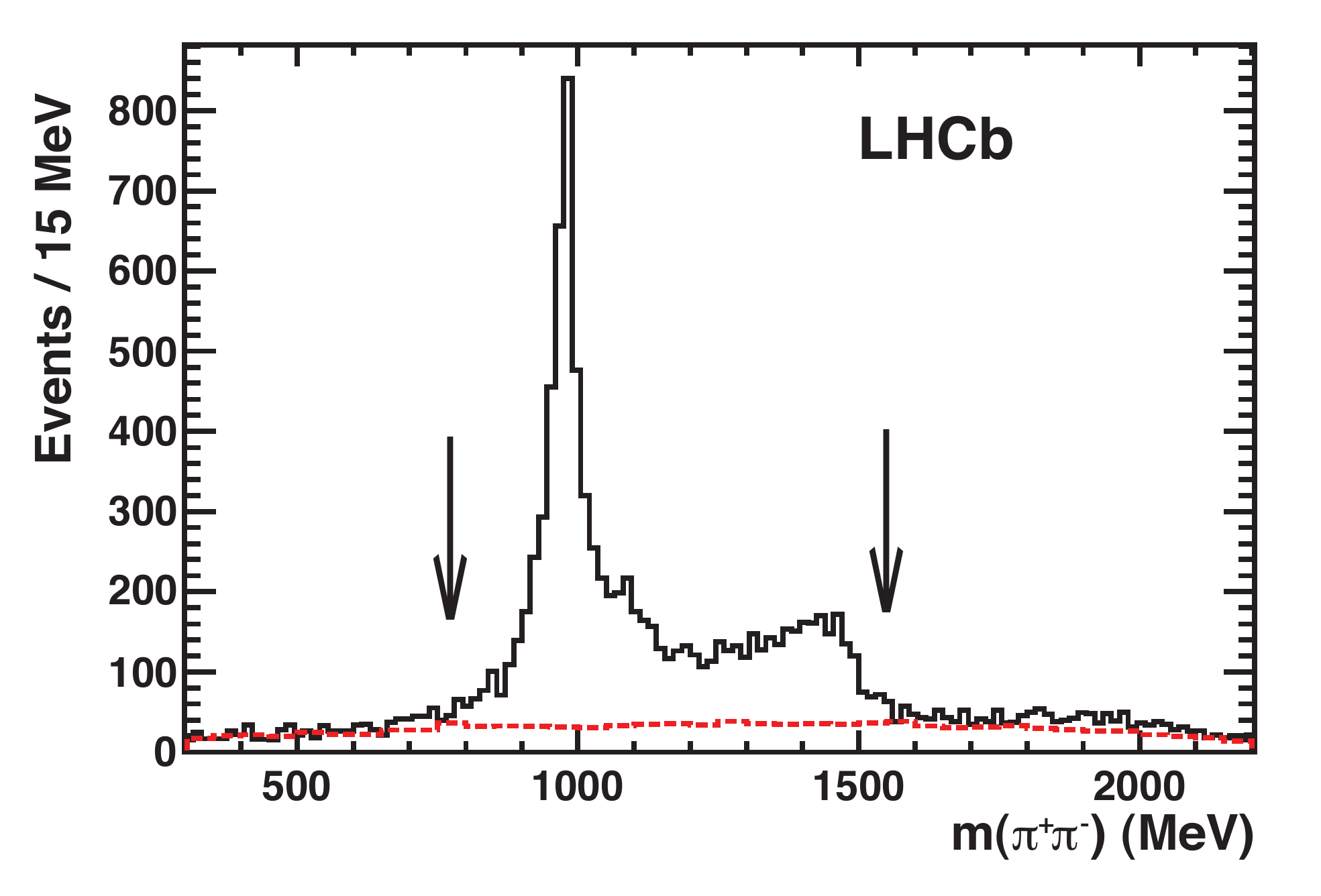}
\vspace{-4mm}
\caption{Mass distribution of selected $\pi^+\pi^-$ combinations shown as the (solid black) histogram for events in the $\Bsb$ signal region. The (dashed red) line shows the background determined by fitting the $\jpsi\pi^+\pi^-$ mass in bins of $\pi^+\pi^-$ mass. The arrows designate the limits of the $f_{\rm odd}$ region.
 } \label{mpipi}
\end{figure}

The $\jpsi\pi^+\pi^-$ mass distribution  is shown in 
Fig.~\ref{fitmass_bdt_sel} for the $f_{\rm odd}$ region.  
In the $\Bsb$ signal region, defined as $\pm$20 MeV around the $\Bsb$ mass peak, there are 7421$\pm$105 signal events, 1717$\pm$38 combinatorial background events, and 66$\pm$9 $\eta'$ background events, corresponding to an 81\% signal purity. The $\pi^+\pi^-$ mass distribution is shown in Fig.~\ref{mpipi}. The most prominent feature is the $f_0(980)$,  containing 52\% of the events within $\pm$90 MeV of 980 MeV, called the $f_0$ region.
 The rest of the $f_{\rm odd}$ region is denoted as $\tilde{f}_0$.

\section{\boldmath Resonance structure in the $\jpsi \pi^+\pi^-$ final state}

The resonance structure in $\Bsb\rightarrow \jpsi \pi^+\pi^-$ decays has been studied using a modified Dalitz plot analysis including the decay angular distribution of the $\jpsi$ meson \cite{LHCb-PAPER-2012-005}. A fit is performed to the decay distributions of several $\pi^+\pi^-$ resonant states described by interfering decay amplitudes. The largest component is the $f_0(980)$ that is described by a Flatt\'e function \cite{Flatte:1976xv}. The data are best described by adding Breit-Wigner amplitudes for the $f_0(1370)$ and $f_2(1270)$ resonances and a non-resonant amplitude.  
The components and fractions of the best fit are given in Table~\ref{tab:ff1}.
\begin{table}[h!]
\begin{center}
\caption{Resonance fractions in $\Bsb\rightarrow \jpsi \pi^+\pi^-$ over the full mass range \cite{LHCb-PAPER-2012-005}. The final-state helicity of the D-wave is denoted by $\Lambda$. Only statistical uncertainties are quoted.}
\begin{tabular}{lcc}
\hline
~~~Resonance &  Normalized fraction (\%)\\
\hline
$f_0(980)$ &$69.7\pm2.3$ \\
$f_0(1370)$ &$21.2\pm2.7$\\
non-resonant $\pi^+\pi^-$ & ~\,$8.4\pm1.5$\\
$f_2(1270)$, $\Lambda=0$ &~\,$0.49\pm0.16$\\
$f_2(1270)$, $|\Lambda| = 1$ &~\,$0.21\pm0.65$\\         
\hline
\end{tabular}
\end{center}\label{tab:ff1}
\vspace{-4mm}
\end{table}

The final state is dominated by \CP-odd S-wave over the entire $f_{\rm odd}$ region. We also have a small D-wave component associated with the $f_2(1270)$ resonance. Its zero helicity ($\Lambda=0$) part  is also pure \CP-odd and corresponds to $(0.49\pm 0.16^{+0.02}_{-0.08})\%$ of the total rate.\footnote{In this Letter whenever two uncertainties are given, the first is statistical and the second systematic.}
 The $|\Lambda|=1$ part, which is of mixed \CP, corresponds to $(0.21\pm 0.65^{+0.01}_{-0.03})$\% of the total. Performing a separate fit, we find that a possible $\rho$ contribution is smaller than 1.5\% at 95\% confidence level (CL). Summing the $f_2(1270)$ $|\Lambda|=1$ and $\rho$ rates, we find that the \CP-odd fraction is larger than 0.977 at 95\% CL. Thus the entire mass range can be used to study \CP violation in this almost pure \CP-odd final state.


\section{Flavour tagging}
\label{sec:tagging}

Knowledge of the initial $\Bsb$ flavour is necessary in order to use Eq.~\ref{eq:CPrate}. This is realized by tagging the
flavour of the other $b$ hadron in the event, exploiting information from four sources: the charges of muons, electrons, kaons with significant IP, and inclusively reconstructed secondary vertices.  The decisions of the four tagging algorithms are individually calibrated using $B^{\mp}\to  \jpsi K^{\mp}$ decays and combined using a neural network as described in Ref.~\cite{Aaij:2012mu}. 
The tagging performance is characterized by $ \varepsilon_{\rm tag} D^2$, where $ \varepsilon_{\rm tag}$ is the efficiency and $D$ the dilution, defined as $D\equiv(1-2\omega)$, where $\omega$ is the probability of an incorrect tagging decision.

We use both the information of the tag decision and of the predicted per-event mistag probability. The calibration procedure assumes a linear dependence between the predicted mistag probability $\eta_i$  for each event and the actual mistag probability $\omega_i$ given by
$\omega_i=p_0+p_1\cdot\left(\eta_i-\langle \eta\rangle \right)$,
where $p_0$ and $p_1$ are calibration parameters and $\langle \eta \rangle$ the average estimated mistag probability as determined from the $\jpsi K^{\mp}$ calibration sample. The values are
$p_0=0.392\pm 0.002\pm 0.009$,  $p_1=1.035\pm0.021\pm0.012$, and  $\langle \eta \rangle=0.391$. Systematic uncertainties are evaluated by using  $\jpsi K^+$  separately from $\jpsi K^-$,  performing the calibration with $\Bzb\to \jpsi\Kstarzb$ and $\Bzb\to D^{*+}\mu^-\overline{\nu}_{\mu}$ plus charge-conjugate channels, and viewing the dependence on different data taking periods. 
We find $ \varepsilon_{\rm tag}=(32.9 \pm0.6)$\% providing us with 2445
tagged signal events.  The dilution is measured as $D=0.272\pm0.004\pm0.015$,
leading to $ \varepsilon_{\rm tag} D^2=(2.43\pm 0.08\pm 0.26)$\%. 

\section{Decay time resolution}
The $\Bsb$ decay time is defined here as
$t = m {\vec{d}\cdot\vec{p}}/{|p|^2}$,
where $m$ is the reconstructed invariant mass, $\vec{p}$ the
momentum and $\vec{d}$ the vector from the primary to the secondary vertex. 
The time resolution for signal increases by about 20\% for decay times from 0 to 10 ps, according to both the simulation and the estimate of the resolution from the reconstruction. 
To take this dependence  into account, we use a double-Gaussian resolution function with widths proportional to the event-by-event estimated resolution,

\begin{equation}
\label{eq:ebyetime}
T(t-\hat{t};\sigma_t)=\sum_{i=1}^2 f^T_i \frac{1}{\sqrt{2\pi}S_i\sigma_t}e^{-\frac{(t-\hat{t}-\mu_t)^2}{2(S_i\sigma_t)^2}}\,,
\end{equation}
where  $\hat{t}$ is the true time, $\sigma_t$ the estimated time resolution, $\mu_t$ is the bias on the time measurement, $f^T_1+f^T_2=1$ are the fractions of each Gaussian, and $S_1$ and $S_2$ are scale factors.  

\begin{figure}[hbt]
\centering
\includegraphics[width=4in]{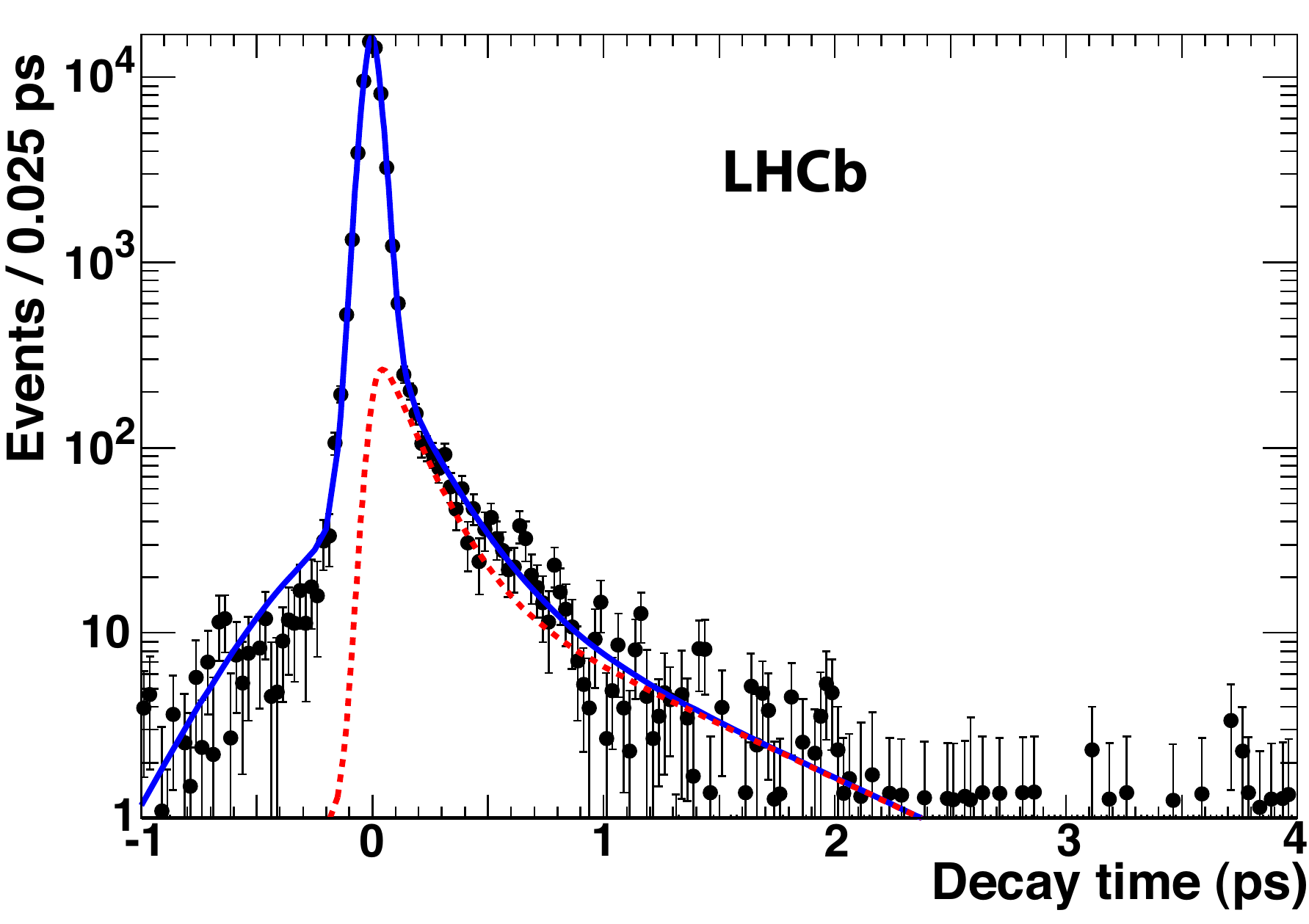}
\caption{Decay time distribution of prompt $\jpsi\pi^+\pi^-$ candidates in the $f_0$ region. The dashed (red) line shows the long-lived component, and the solid curve the total.} \label{timeresDATA}
\end{figure}

To determine the parameters of $T$ we use
events containing a $\jpsi$, found using a dimuon trigger without track impact parameter requirements, plus two opposite-sign charged tracks with similar selection criteria as for $\jpsi \pi^+\pi^-$ events including that the  $\jpsi\pi^+\pi^-$ mass be within $\pm$20 MeV of the $\Bsb$ mass.
 Figure~\ref{timeresDATA} shows the decay time distribution for this $\jpsi\pi^+\pi^-$ prompt data sample for the $f_0$ region; the $\tilde{f}_0$ data are very similar. The data are fitted with the time dependence given by
 \begin{equation}
\label{eq:prompttime}
P^{\rm prompt}(t)=(1-f_1-f_2) T(t;\sigma_t)+\left[\frac{ f_1}{\tau_1}e^{{-\hat{t}}/{\tau_1}}+\frac{ f_2}{\tau_2}e^{{-\hat{t}}/{\tau_2}}\right]\otimes T(t-\hat{t};\sigma_t)\,,
\end{equation}
where $f_1$ and $f_2$ are long-lived background fractions with lifetimes $\tau_1$ and $\tau_2$, respectively.
 The resulting parameter values of the function $T$ are given in Table~\ref{tab:timerest}.
\begin{table}[h!]
\begin{center}
\caption{Parameters of the decay time resolution function determined from fits to $\jpsi\pi^+\pi^-$ prompt data samples.}
\begin{tabular}{ccc}
\hline
Parameter & $f_0$ region & $\tilde{f}_0^{\Large\vphantom{X}}$ region\\
\hline
$\mu_t$ (fs) & $-$3.32(12) & $-$2.91(7)\\
$S_1$ & 1.362(4) & 1.329(2)\\
$S_2$ & 12.969(3) & 9.108(3)\\
$f^T_2$ & 0.0193(7) & 0.0226(5)\\       
\hline
\end{tabular}
\end{center}\label{tab:timerest}
\end{table}

\begin{figure}[hbt]
\centering
\includegraphics[width=4.3in]{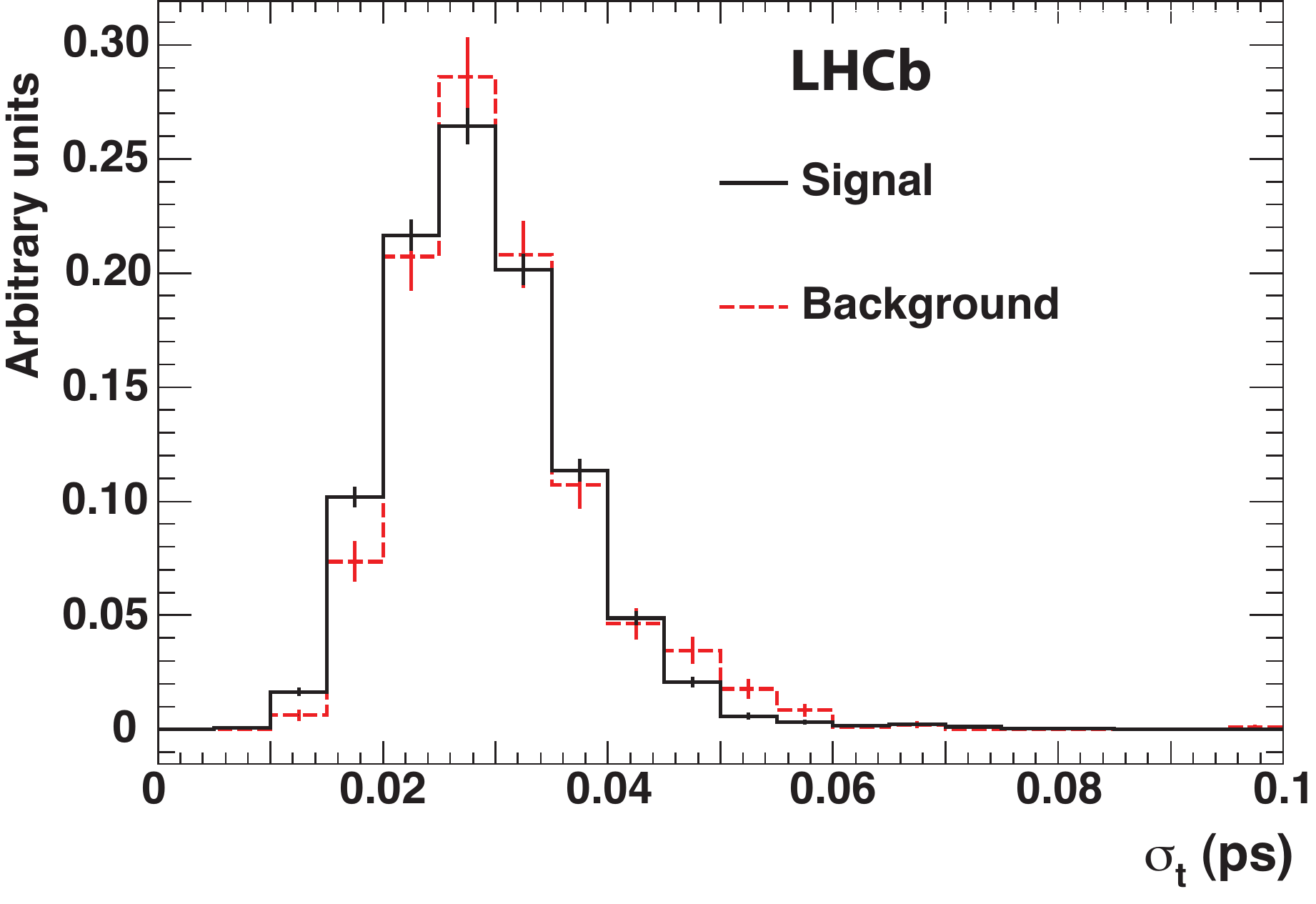}
\vspace{-4mm}
\caption{Distribution of the estimated time resolution $\sigma_t$ for opposite-sign $\jpsi \pi^+\pi^-$ signal events after background subtraction, and for like-sign background.}
\label{cmp}
\end{figure}

Figure~\ref{cmp} shows the
 $\sigma_t$ distributions  used in Eq.~\ref{eq:ebyetime}  for  $\jpsi\pi^+\pi^-$ events in the $f_{\rm odd}$ region after background subtraction, and for like-sign background.  
 Taking into account the calibration parameters of Table~\ref{tab:timerest}, the average effective decay time resolution for the signal is  40.2 fs and 39.3 fs for the $f_0$  and $\tilde{f}_0$ regions, respectively. The average of the two samples is 39.8 fs.

\section{Decay time acceptance}

\begin{figure}[hbt]
\centering
\includegraphics[width=6.in]{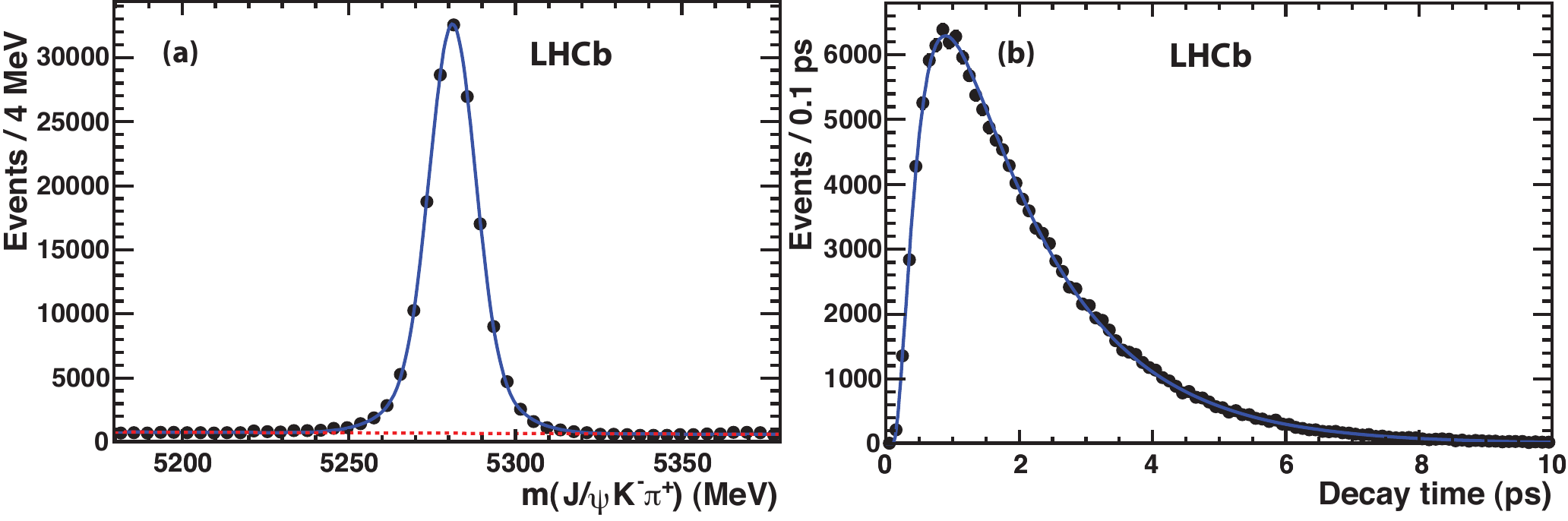}
\caption{(a) Mass distribution of  $\Bzb\to \jpsi \Kstarzb$ candidates. The dashed (red) line shows the background, and the solid (blue) curve the total.  (b) Decay time distribution, where the small background has been subtracted using the $\Bzb$ mass sidebands. The (blue) curve shows the lifetime fit.} \label{mass-psikst}
\end{figure}

The decay time acceptance function is written as
\begin{equation}
\label{eq:accept}
A(t;a,n,t_0)=C\frac{\left[a\left(t-t_0\right)\right]^n}{1+\left[a\left(t-t_0\right)\right]^n}\,,
\end{equation}
where $C$ is a normalization constant. The other parameters are determined by fitting the lifetime distribution of $\Bzb\to \jpsi\Kstarzb$ events, where $\Kstarzb\to K^-\pi^+$. Figure~\ref{mass-psikst}(a) shows the \jpsi\Kstarzb mass when the $K^-\pi^+$ invariant mass is within $\pm$300 MeV of 892 MeV. There are 155,743$\pm$434 signal events. The sideband-subtracted decay time distribution is shown in Fig.~\ref{mass-psikst}(b) together with a lifetime fit taking into account the acceptance and resolution. This fit yields $a=2.11\pm 0.04$\,ps$^{-1}$, $n=1.82\pm 0.06$, $t_0=0.105\pm 0.006$\,ps and a lifetime of 1.516$\pm$0.008\,ps, in good agreement with the PDG average of 1.519$\pm$0.007\,ps \cite{PDG}.

\begin{figure}[hbt]
\centering
\includegraphics[width=4.in]{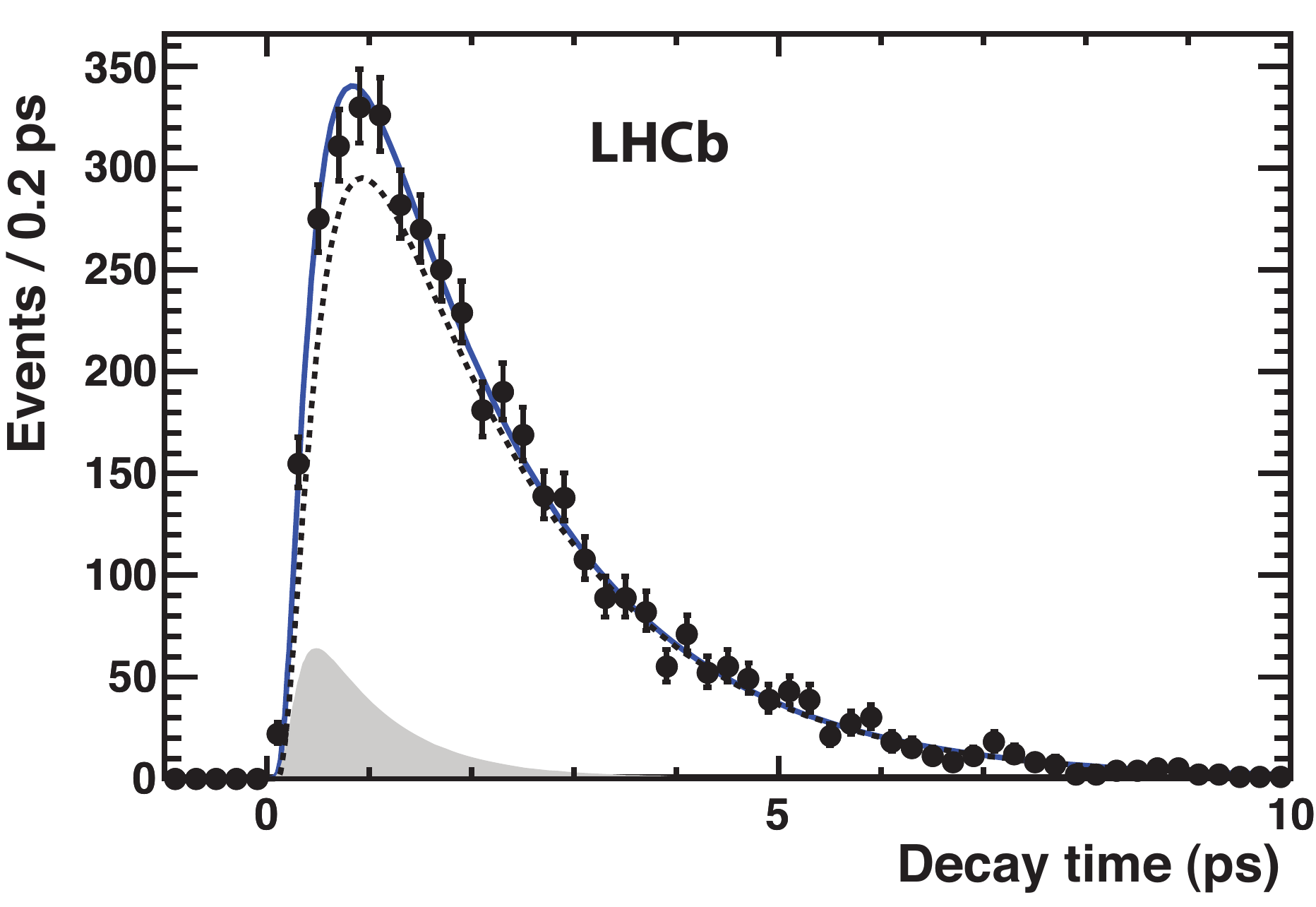}
\vskip -4mm
\caption{Decay time distribution of $\Bsb\to \jpsi f_0$ candidates fitted with a single exponential function multiplied by the acceptance and convolved with the resolution. The dashed line is signal and the shaded area background.} \label{fitlifetime}
\end{figure}

We check our lifetime acceptance by comparing with a CDF measurement of the $\Bsb\to \jpsi f_0$ effective lifetime of $\tau^{\rm eff}=1.70^{+0.12}_{-0.11}\pm 0.03$\,ps \cite{Aaltonen:2011nk} obtained from a single exponential fit.\footnote{This corresponds to the lifetime of the \CP-odd eigenstate if $\phi_s$ is zero (see Eq.~\ref{eq:untagged}).}  
A fit of the $f_0$ sample (see Fig.~\ref{fitlifetime}) yields $\tau^{\rm eff}=1.71\pm 0.03$\,ps, while we find $\tau^{\rm eff}=1.67\pm 0.03$\,ps in the  $\tilde{f}_0$  sample.  These two values are consistent with each other, within the quoted statistical errors, and with the CDF result.

\section{Likelihood function definition}

To determine $\phi_s$  an extended likelihood function is maximized using candidates in the \Bsb signal region
\begin{equation}
\label{eq:like}
{\cal L}(\phi_s)=e^{-(N_{\rm sig}+N_{\rm bkg})}\prod_{i=1}^{N_{\rm obs}} P(m_i,t_i,{\sigma_t}_i, q_i,\eta_i)\,,
\end{equation} 
where the signal yield, $N_{\rm sig}$, and background yield, $N_{\rm bkg}$, are fixed from the fit of the $\jpsi\pi^+\pi^-$ mass distribution in the $f_{\rm odd}$ region  (see Fig.~\ref{fitmass_bdt_sel}). 
$N_{\rm obs}$ is the number of  $\Bsb$ candidates, $m_i$ the reconstructed mass, $t_i$ the reconstructed decay time, and ${\sigma_t}_i$ the estimated decay time uncertainty. The flavour tag, $q_i$, takes values of +1, $-1$ or 0, respectively, if the signal meson is tagged as $B_s^0$,  $\Bsb$, or untagged, and $\eta_i$ is the estimated mistag probability.  Backgrounds are caused largely by mis-reconstructed $b$-hadron decays, so it is necessary to include a long-lived background probability density function (PDF). 
The likelihood function includes distinct contributions from
the signal and the background. 
For tagged events we have 
\begin{eqnarray}
\label{eq:taggedPDF}
P(m_i, t_i, {\sigma_t}_i, q_i,\eta_i)&=&N_{\rm sig}  \varepsilon_{\rm tag} P_m^{\rm sig}(m_i) P_t^{\rm sig}(t_i,q_i,\eta_i \vert{\sigma_t}_i)P_{\sigma_t}^{\rm sig}({\sigma_t}_i)\nonumber\\
&&+N_{\rm bkg}\varepsilon^{\rm bkg}_{\rm tag}P_m^{\rm bkg}(m_i)P_t^{\rm bkg}(t_i\vert{\sigma_t}_i)P_{\sigma_t}^{\rm bkg}({\sigma_t}_i)\,,
\end{eqnarray}
where $\varepsilon^{\rm bkg}_{\rm tag}$ refers to the flavour tagging efficiency of the background.   The signal mass PDF, $P_m^{\rm sig}(m)$, is a double Gaussian function, while the background mass PDF, $P_m^{\rm bkg}(m)$, is proportional to $e^{-\alpha m}$ together with a very small contribution from $\Bsb\to \jpsi\eta'$, $N_{\eta'}$, that is fixed in the $\phi_s$ fit to 66 events obtained from the fit shown in Fig.~\ref{fitmass_bdt_sel}. 

The PDF used to describe the signal decay rate, $P_t^{\rm sig}$, depends on the tagging results $q$ and $\eta$. It is modelled
by a PDF of the true time $\hat{t}$, $R(\hat{t}, q, \eta)$, convolved with the decay time resolution
 and  multiplied
by the decay time acceptance function found for $\Bzb\to \jpsi \Kstarzb$ events.
 From Eq.~\ref{eq:CPrate}, it can be expressed as
\begin{equation}
R(\hat{t},q,\eta) \propto e^{-\Gamma_s \hat{t}}\left\{\cosh\frac{\Delta\Gamma_s
\hat{t}}{2}+\cos\phi_s\sinh\frac{\Delta\Gamma_s \hat{t}}{2}- q [1-2\omega(\eta)]
\sin\phi_s\sin(\Delta m_s \hat{t})\right\}\,,\label{eq-sigt}
\end{equation}
where $\omega(\eta)$ is the calibrated mistag probability.
Thus the PDF of reconstructed time is
\begin{equation}
P_t^{\rm sig}(t,q,\eta\vert{\sigma_t})=R(\hat{t},q,\eta)\otimes T(t-\hat{t};\sigma_t)\cdot A(t^{\rm};a,n,t_0)\,.
\end{equation}
For untagged events we use
\begin{eqnarray}
\label{eq:untaggedPDF}
P(m_i,t_i,{\sigma_t}_i,q_i=0,\eta_i)& = &N_{\rm sig}(1-  \varepsilon_{\rm tag})P_m^{\rm sig}(m_i)P_t^{\rm sig}(t_i,0,\eta_i\vert{\sigma_t}_i)P_{\sigma_t}^{\rm sig}({\sigma_t}_i) \nonumber\\
&&+ N_{\rm bkg}(1-\varepsilon^{\rm bkg}_{\rm tag})P_m^{\rm bkg}(m_i)P_t^{\rm bkg}(t_i\vert{\sigma_t}_i) P_{\sigma_t}^{\rm bkg}({\sigma_t}_i)\,.
\end{eqnarray}

The PDF describing the long-lived background decay rate is
\begin{equation}
P^{\rm bkg}_t(t\vert\sigma_t)=\left[\frac{1-f^{\rm bkg}_2}{\tau^{\rm bkg}_1}e^{-\frac{\hat{t}}{\tau^{\rm bkg}_1}}+\frac{f^{\rm bkg}_2}{\tau^{\rm bkg}_2}
e^{-\frac{\hat{t}}{\tau^{\rm bkg}_2}}\right]\otimes
T(t^{\rm}-\hat{t};\sigma_t)\cdot
A(t^{\rm};a^{\rm bkg},n^{\rm bkg},t_0^{\rm bkg})\,,
\end{equation}
where $\tau^{\rm bkg}_1$, $\tau^{\rm bkg}_2$ and $f^{\rm bkg}_2$ parameterize the underlying double exponential function.
The same functional form is used to describe the background decay time acceptance as for signal (Eq.~\ref{eq:accept}) with different parameters that are determined by fitting the like-sign $\jpsi\pi^{\pm}\pi^{\pm}$ events in an interval $\pm$200 MeV around the $\Bsb$ mass.
The $P^{\rm sig}_{\sigma_t}({\sigma_t}_i)$  and $P^{\rm bkg}_{\sigma_t}({\sigma_t}_i)$ functions are shown in Fig.~\ref{cmp}.
The parameters that are fixed in the likelihood fit are listed in Table~\ref{tab:PDFs}.

\begin{table}[htb]
\center
\caption{\label{tab:PDFs} 
Parameters used in the functions for the invariant mass and decay
time describing the signal and background. These parameters are fixed to their central values
in the fit for $\phi_s$.
}  
\begin{tabular}{ll}\hline
{Function} & { Parameters}\\\hline
& $N_{\rm sig}=7421$, $N_{\rm bkg}=1717\pm 38$, $N_{\eta'}=66\pm 9$\\
$P_m^{\rm sig}(m)$ & $m_0$= 5368.2(1) MeV, $\sigma^m_1$=8.1(1) MeV, $\sigma^m_2$=18.0(2) MeV, $f^m_2 $= 0.196(2)\\
$P_m^{\rm bkg}(m)$& $\alpha=(-5.35\pm1.15)\times 10^{-4}$\,MeV$^{-1}$\\
$P^{\rm bkg}_t(t\vert\sigma_t)$ & $\tau^{\rm bkg}_1=0.65(5)$\,ps, $\tau^{\rm bkg}_2=2.0(8)$\,ps, $f^{\rm bkg}_2=0.06(2)$  \\
&  $a^{\rm bkg}=3.22(10)$ ps$^{-1}$, $n^{\rm bkg}=3.31(14)$, $t_0^{\rm bkg}=0$\,ps,\\
$T(t-\hat{t}; \sigma_t)$ & see Table~\ref{tab:timerest}\\
\hline
\end{tabular}
\end{table}

\section{Results}
The likelihood of Eq.~\ref{eq:like} is multiplied by Gaussian constraints on several of the model parameters.
These are the
 LHCb measured value of $\Delta m_s =17.63\pm 0.11\pm 0.02$\,ps$^{-1}$ \cite{Aaij:2011qx}, the tagging parameters $p_0$ and $p_1$, the decay time acceptance parameters $t_0$, $a$, and $n$, and both $\Gamma_s=0.657\pm0.009\pm0.008$\,ps$^{-1}$ and  $\Delta\Gamma_s=0.123\pm 0.029\pm 0.011$\,ps$^{-1}$ given by the $\jpsi\phi$ analysis  \cite{LHCb:2011aa}. The fit has been validated with full Monte Carlo simulations.
  
\begin{figure}[bt]
\centering
\includegraphics[width=4.in]{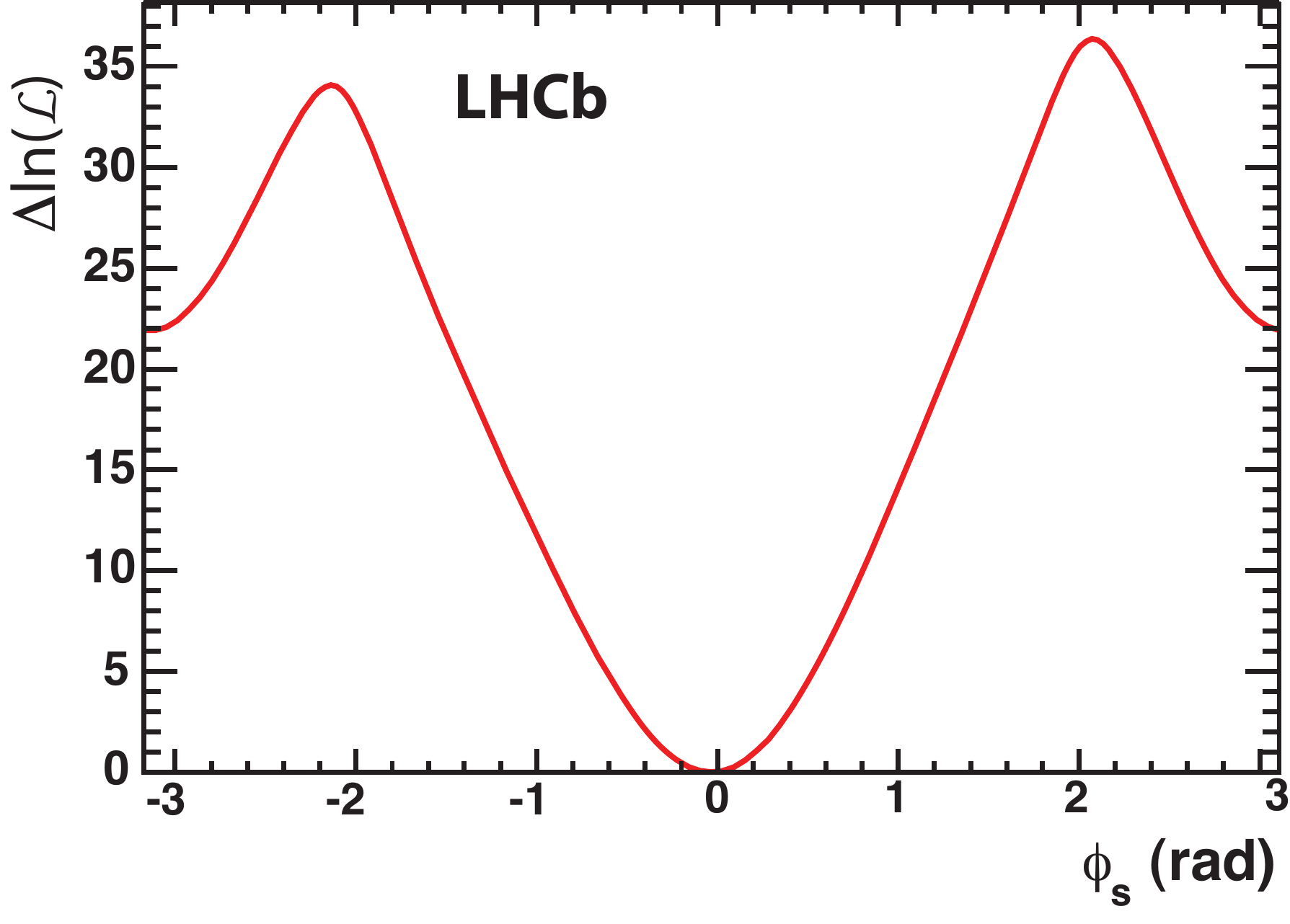}
\vspace*{-4mm}
\caption{Log-likelihood difference as a function of  $\phi_s$  for  $\Bsb\to \jpsi f_{\rm odd}$ events. } \label{dll-new}
\vspace*{4mm}
\end{figure}

\begin{figure}[hbt]
\centering
\hspace*{6mm}\includegraphics[width=4in]{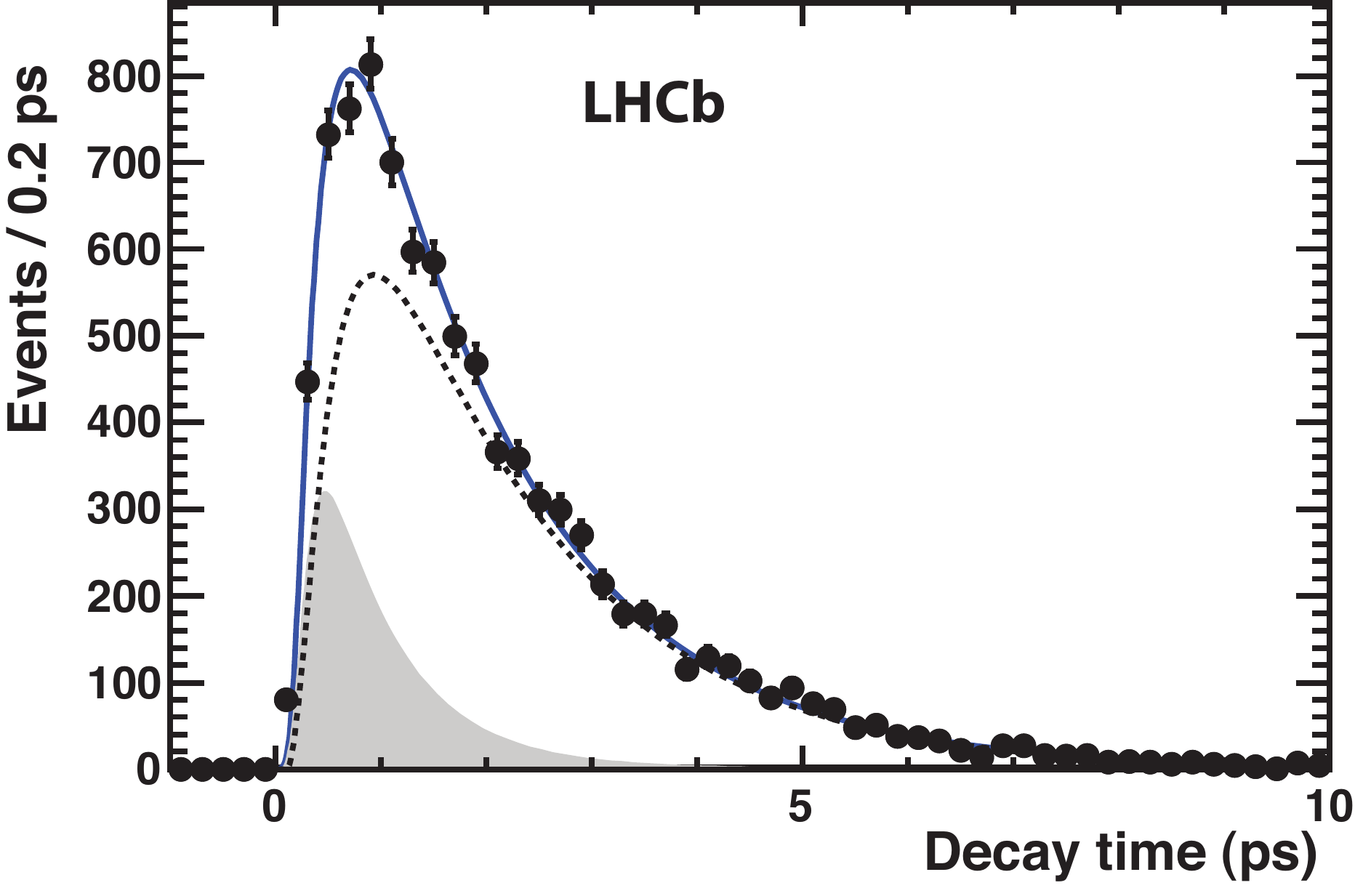}
\vspace{-2mm}
\caption{ Decay time distribution of $\Bsb\to \jpsi f_{\rm odd}$ candidates. The solid line shows the result of the fit, the dashed line shows the signal, and
the shaded region the background.}
 \label{time2}
\end{figure}

Figure~\ref{dll-new} shows the difference of log-likelihood value, $\Delta \ln(\cal{L})$, compared to the one at the point with the best fit, as a function of $\phi_s$. At each value, the likelihood function is maximized with respect to all other parameters.  
The best fit value is $\phi_s=-0.019^{+0.173+0.004}_{-0.174-0.003}$\,rad.  (The systematic uncertainty will be discussed subsequently.) Values for $\phi_s$ in the $f_0$ and $\tilde{f}_0$ regions  are $-0.26\pm0.23$\,rad and $0.29\pm0.28$\,rad, respectively, consistent within the uncertainties.
The decay time distribution is shown in Fig.~\ref{time2}. 

\begin{figure}[hbt]
\centering
\includegraphics[width=4.in]{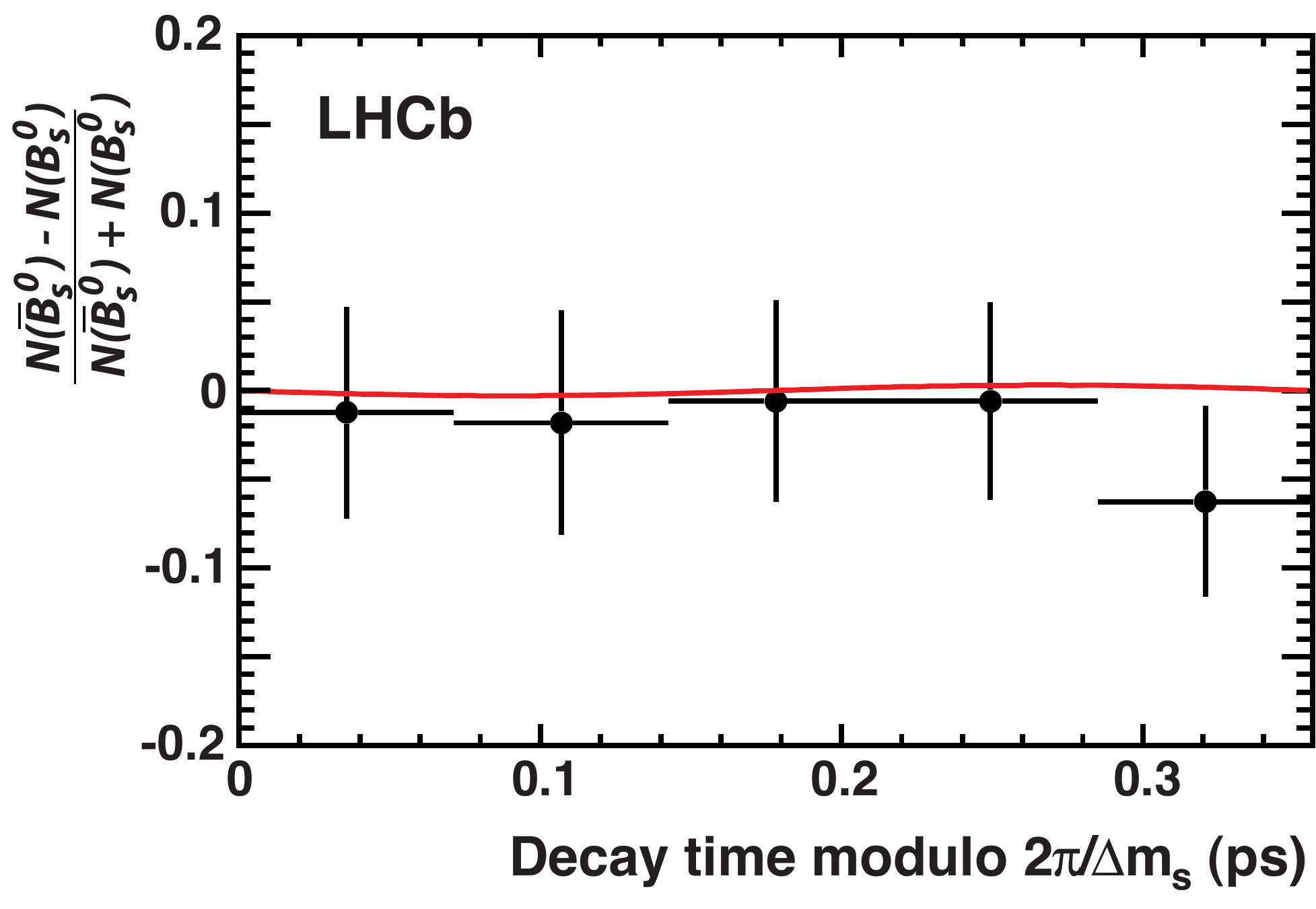}
\caption{\CP asymmetry as a function of decay time modulo $2\pi/\Delta m_s$. The curve shows the expectation for $\phi_s=-0.019$\, rad. } \label{asym_modu}
\vspace{-2mm}
\end{figure}

 The presence of a $\sin\phi_s$  contribution in Eq.~\ref{eq:CPrate}  can, in principle, be viewed by plotting the asymmetry $\left[ N( \Bsb)- N( \Bs)\right]/\left[N ( \Bsb)+ N ( \Bs)\right]$ of the background-subtracted tagged yields as a function of decay time modulo $2\pi/\Delta m_s$, as shown in Fig.~\ref{asym_modu}. The asymmetry is consistent with the value of $\phi_s$ determined from the full fit and does not show any significant structure.

The data have also been analyzed allowing for the possibility of direct \CP violation.
In this case Eq.~\ref{eq-sigt} must be replaced with
\begin{eqnarray}
R(\hat{t},q,\eta) &\propto& e^{-\Gamma_s \hat{t}}\left\{\cosh\frac{\Delta\Gamma_s
\hat{t}}{2}+\frac{2|\lambda|}{1+|\lambda|^2}\cos\phi_s\sinh\frac{\Delta\Gamma_s \hat{t}}{2}\right. \nonumber\\
& &
\left. -\frac{q[1-2\omega(\eta)]}{1+|\lambda|^2}\left[2|\lambda|\sin\phi_s\sin(\Delta m_s \hat{t})
-(1-|\lambda|^2)\cos(\Delta m_s\hat{t})\right]
\right\}\,.\label{eq-sigt2}
\end{eqnarray}
The fit gives $ |\lambda|=0.89\pm 0.13$, consistent with no direct \CP violation ($|\lambda|=1$). The value
of $\phi_s$ changes only by $-0.002$\, rad, and the uncertainty stays the same.


The systematic uncertainties are small compared to the statistical one. No additional uncertainty is introduced by the acceptance parameters, $\Delta m_s$, $\Gamma_s$, $\Delta\Gamma_s$ or flavour tagging, since Gaussian constraints are applied in the fit.  The uncertainties associated with the fixed parameters are evaluated by changing them by $\pm$1 standard deviation from their nominal values and determining the change in the fitted value of $\phi_s$. These are listed in Table~\ref{tab:syserr}. The uncertainty due to a change in the signal time acceptance function is evaluated by
multiplying $A(t;a,n,t_0)$ with a factor
$(1+\beta t)$, and redoing the $\Bzb\to\jpsi\Kstarzb$ fit with the $\Bzb$ lifetime fixed to the PDG value. The resulting value of $\beta=(1\pm 3\pm 3)\times 10^{-3}$ is then varied by $\pm 4.4\times 10^{-3}$ to estimate the uncertainty in $\phi_s$.
An additional uncertainty is included due to a possible  \CP-even component. This has been limited to 2.3\% of the total $f_{\rm odd}$ rate at 95\% CL, and contributes an uncertainty to $\phi_s$ as determined by repeating the fit with an additional multiplicative dilution of 0.954. The asymmetry between $B^0_s$ and $\Bsb$ production is believed to be small, and similar to the asymmetry between $B^0$ and $\Bzb$ production which has been measured by LHCb to be about 1\% \cite{Aaij:2012qe}. The effect of neglecting this production asymmetry is the same as making a relative 1\% change in the tagging efficiencies, up for \Bs and down for \Bsb, which has a negligible effect on $\phi_s$.
\begin{table}[!htb]
\centering
\caption{Summary of systematic uncertainties on $\phi_s$. Quantities fixed in the fit that are not included here give negligible uncertainties. The total uncertainty is found by adding in quadrature all the positive and negative contributions separately.}
\label{tab:syserr}
\begin{tabular}{ccrr}
\hline
Quantity (Q) & $\pm\Delta$Q & $+$Change~~~& $-$Change~~~\\
              &                                    &in $\phi_s$\,(rad) & in $\phi_s$\,(rad)\\
\hline
$\beta$ & $4.4\times 10^{-3}$ &0.0008~~~&$-0.0007$~~~\\
$\tau^{\rm bkg}_1$ (ps) &0.046&$-0.0006$~~~&0.0014~~~\\
$\tau^{\rm bkg}_2$ (ps) &0.8&$-0.0014$~~~&0.0014~~~\\
$f^{\rm bkg}_2$ & 0.02 & $-0.0006$~~~& 0.0012~~~\\
$N_{\rm bkg}$  & 38 &0.0009~~~& $-0.0001$~~~\\
$N_{\eta'}$ & 9	& 0.0006~~~& 0.0001~~~\\	
$m_0$ (MeV)& 0.12& 0.0012~~~& $-0.0004$~~~\\
$\sigma^m_1$ (MeV)& 0.1& $-0.0002$~~~&0.0008~~~\\
$\alpha$ &$1.1\times 10^{-4}$&$0.0003$~~~&$0.0003$~~~\\
$T$ function &5\%& $0.0005$~~~&0.0005~~~\\
\CP-even  & multiply dilution by 0.954& $-0.0008$~~~&$-$~~~~~~\\
Direct \CP & free in fit & $-0.0020$~~~&$-$~~~~~~\\
\hline
\multicolumn{2}{l}{Total systematic uncertainty on $\phi_s$}&\multicolumn{2}{c}{$^{+0.004}_{-0.003}$}\\
\hline
\end{tabular}
\end{table}

\section{Conclusions} 
Using 1 fb$^{-1}$ of data collected with the LHCb detector, 
$\Bsb\to \jpsi \pi^+\pi^-$ decays are selected and used to measure the \CP violating phase $\phi_s$. 
The signal events have an effective decay time resolution of 39.8 fs. The flavour tagging is based on properties of the decay of the other $b$ hadron in the event and has an efficiency times dilution-squared of 2.4\%. We perform a fit of the time dependent rates with the $\Bsb$ lifetime  and the difference in widths of the heavy and light eigenstates used as input. We measure a value of $\phi_s= -0.019^{+0.173+0.004}_{-0.174-0.003}$\,rad. This result subsumes our previous measurement obtained with 0.41\,\invfb of data \cite{LHCb:2011ab}.
Combining this result with our previous result from $\Bsb\to \jpsi\phi$ decays \cite{LHCb:2011aa} by performing a joint fit to the data gives a combined LHCb value of $\phi_s= +0.06\pm 0.12\pm0.06$\,rad. Our result is consistent with the SM prediction of $-0.0363^{+0.0016} _{-0.0015}$\,rad  \cite{Charles:2011va}.  In addition, we find no evidence for direct \CP violation.

\section*{Acknowledgements}
\noindent We express our gratitude to our colleagues in the CERN accelerator
departments for the excellent performance of the LHC. We thank the
technical and administrative staff at CERN and at the LHCb institutes,
and acknowledge support from the National Agencies: CAPES, CNPq,
FAPERJ and FINEP (Brazil); CERN; NSFC (China); CNRS/IN2P3 (France);
BMBF, DFG, HGF and MPG (Germany); SFI (Ireland); INFN (Italy); FOM and
NWO (The Netherlands); SCSR (Poland); ANCS (Romania); MinES of Russia and
Rosatom (Russia); MICINN, XuntaGal and GENCAT (Spain); SNSF and SER
(Switzerland); NAS Ukraine (Ukraine); STFC (United Kingdom); NSF
(USA). We also acknowledge the support received from the ERC under FP7
and the Region Auvergne.

\ifx\mcitethebibliography\mciteundefinedmacro
\PackageError{LHCb.bst}{mciteplus.sty has not been loaded}
{This bibstyle requires the use of the mciteplus package.}\fi
\providecommand{\href}[2]{#2}


\end{document}